\documentclass[prl,aps,twocolumn,superscriptaddress,amsmath,amssymb,floatfix,longbibliography]{revtex4-2}

\usepackage{graphicx}
\usepackage{bm}
\usepackage{hyperref}
\usepackage{xcolor}
\usepackage{soul}

\begin{document}

%\title{Stochastic resetting search with correlated stochastic target}

\title{Optimal Evasion from a Resetting Searcher}

\author{Ami Taitelbaum}
\affiliation{Racah Institute of Physics, The Hebrew University of Jerusalem, Jerusalem 9190401, Israel}
\author{Matan Shporer}
\affiliation{Racah Institute of Physics, The Hebrew University of Jerusalem, Jerusalem 9190401, Israel}
\author{Shlomi Reuveni}
\affiliation{School of Chemistry, The Center for Computational Molecular and Materials Science, The Center for Physics and Chemistry of Living Systems, Tel Aviv University, Tel Aviv, Israel}
\author{Michael Assaf}
\affiliation{Racah Institute of Physics, The Hebrew University of Jerusalem, Jerusalem 9190401, Israel}

%\date{\today}

\begin{abstract}
Returning to a familiar location can help a searcher find its target—but how should the target move to avoid being found? We address this question for a resetting Brownian searcher and a mobile target fluctuating around a home location. Keeping the target’s typical exploration range fixed, we find that the mean capture time is maximized at an intermediate correlation time: rapid motion promotes encounters, whereas slow motion leaves the target effectively stationary. Between these extremes, persistent excursions away from the searcher’s reset point can substantially delay capture. Analytical limiting results explain this optimum, while numerical solutions reveal an approximately linear relation between the maximal mean capture time and the optimal correlation time across many decades. These findings identify temporal correlations as a resource for evasion, allowing a target to prolong survival by changing how it moves rather than where it roams.
\end{abstract}

\maketitle

\textit{Introduction.} Stochastic resetting, the act of taking a stochastic process and starting it anew \cite{EvansMajumdarSchehr2020,gupta2022stochastic,nagar2023stochastic,kundu2024preface,keidar2026stochastic}, is often studied from the perspective of a searcher looking for a target \cite{Kusmierz2014,bhat2016stochastic,pal2017first,chechkin2018random,mercado2018lotka,pal2019first,yin2023restart,pal2024random}. For example, by repeatedly returning to a prescribed location, a diffusive searcher can avoid unproductive excursions, acquire a finite mean first-passage time, and reach a target faster~\cite{EvansMajumdar2011,tal2020experimental,besga2020optimal}. This idea has been extended to a broad range of stochastic search and completion processes~\cite{luby1993optimal,montanari2002optimizing,Reuveni_Role_2014,Reuveni2016,roldan2016stochastic,pal2017first,bonomo2022mitigating,blumer2024combining,blumer2025have}, and the resulting optimization problem is  well understood when the target, or its distribution, are fixed \cite{EvansMajumdar2011Optimal,campos2015phase,Reuveni2016,boyer2017long,pal2019landau,ray2019peclet,rotbart2015michaelis,vilk2022fluctuations,evans2025stochastic}.

Here we reverse the searcher-centric perspective and ask instead: if the target is mobile and seeks to delay capture, how should it move?
First-encounter problems with mobile targets have been studied in diffusive predator-prey systems~\cite{RednerKrapivsky1999,gabel2012can,EvansMajumdarSchehr2022}. While in broad classes of classical trapping models unbiased target motion cannot improve survival relative to remaining immobile~\cite{MoreauEtAl2003Pascal},  reactive evasion can provide a marked survival advantage~\cite{OshaninEtAl2009}. A resetting searcher opens a third possibility: the target can counter the searcher's capture strategy without sensing or reacting to its motion, solely by tuning the temporal organization of its autonomous dynamics. We introduce this target-side optimization problem for a target fluctuating within a fixed spatial range around a preferred ``home'' location and ask: how rapidly should it explore that range to evade capture for as long as possible?
%\st{Here, we take a complementary but fundamentally different perspective---the target follows autonomous stochastic dynamics and neither senses nor reacts to the searcher---and ask how the target should
%move to optimally evade a resetting searcher.
%Essentially, we are interested in reversing the classical perspective of resetting problems and ask: 
%i.e., if the target is mobile and seeks to delay capture, how should it move? 
%Concretely, for a target fluctuating within a fixed spatial  range around a preferred ``home" location, how rapidly should it explore that range to evade a resetting searcher for as long as possible?}}

The answer is not obvious. A slowly moving target remains near its initial position for long periods and can therefore be located through repeated search excursions. Increasing its mobility allows it to move away, but motion that is too rapid also causes it to repeatedly revisit the neighborhood of the searcher's reset point, creating opportunities for capture. This suggests a competition between mobility and persistence: effective evasion requires the target to travel sufficiently far from home, while also remaining there long enough to frustrate successive search attempts. The relevant control parameter is therefore not only the spatial range explored by the target, but also the correlation time of its motion.

\begin{figure}[t]
\centering
\includegraphics[width=0.64\columnwidth]{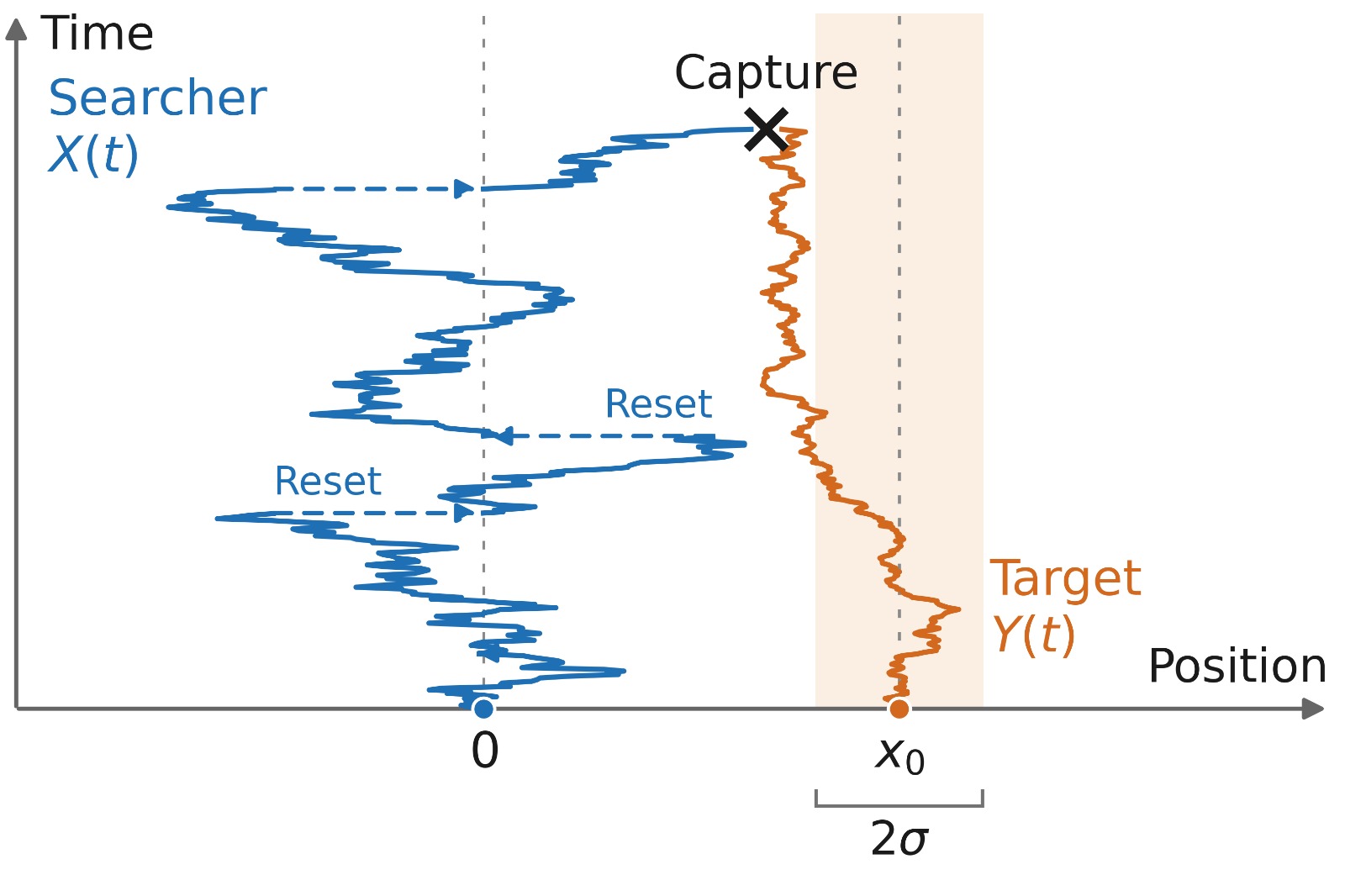}
\vspace{-0.6cm}
\caption{Space-time schematic of a Brownian searcher  that repeatedly resets to the origin, and a target undergoing OU motion around its home at $x_0$. The shaded region indicates the target's typical exploration range, $\sigma$, while the correlation time $\tau$ controls how long excursions away from home persist. Capture occurs when the two trajectories first meet. %Here $D=1$, $r=0.2$, $x_0=6$, $\sigma=1.2$ and $\tau=20$, see model below. %\textcolor{red}{SR: Compare this Fig. to Fig. 2. I would make all fonts and line thickness as is Fig. 2. Here everything is too small/thin. To save space, I would make time (which is the long axis), the x-axis. This way we will exploit the full width of the column while taking less hight.}
}
\label{fig1}
\end{figure}

To address this question, we introduce the minimal model illustrated in Fig.~\ref{fig1}. A Brownian searcher with diffusion coefficient $D$ resets to the origin at rate $r$. The target, in contrast, undergoes Ornstein--Uhlenbeck (OU) motion around a preferred location $x_0$, with stationary variance $\sigma^2$ and correlation time $\tau$. Thus, $\sigma$ sets the typical spatial range explored around home, whereas $\tau$ controls how rapidly this range is explored and how long excursions away from home persist. %Varying $\tau$ at fixed $\sigma$ therefore allows us to isolate the role of temporal persistence without changing the target's stationary spatial distribution. Both particles initially occupy their preferred positions, $X(0)=0$ and $Y(0)=x_0$, and capture occurs when their continuously evolving trajectories first meet.

Although the model is simple, its first-encounter problem cannot be solved using standard stochastic resetting theory. For a fixed target, every reset returns the system to the same state, making successive search excursions statistically independent and allowing the first-passage problem to be solved via a renewal approach~\cite{pal2017first,chechkin2018random,EvansMajumdarSchehr2020}. Here, a reset returns only the searcher: the joint state changes from $(X,Y)$ to $(0,Y)$, while the target retains its current position. Thus, successive search excursions are correlated through the target coordinate, and the usual renewal construction does not close. Yet, the two extreme correlation-time regimes admit analytical treatment and reveal distinct, competing mechanisms of capture.

For very short $\tau$, the target rapidly samples its spatial distribution and repeatedly approaches the searcher's reset point, so the mean capture time (MCT) vanishes linearly  with $\tau$. In the opposite limit, the target is effectively stationary near its home, and as $\tau$ increases the MCT approaches the fixed-target result from above. These opposing asymptotic trends produce a maximum at an intermediate correlation time, where the target can move substantially away from home yet remain there long enough to frustrate successive search attempts. We derive the MCT analytically in the short- and long-correlation time regimes and use their crossover to predict the optimal correlation time and its dependence on the resetting rate and exploration range.

\textit{Model.}
We consider a diffusive searcher, with diffusion constant $D$ and variance $2Dt$, whose position $X(t)$ obeys
\begin{equation}
 \dot{X}(t)=\sqrt{2D}\,\eta_X(t)
\label{eq:Xlangevin}
\end{equation}
between stochastic resetting events, at which $X(t)\longrightarrow 0$ at rate $r$. The assumption here is that once resetting occurs, it moves the searcher instantaneously to the origin. 
The target position $Y(t)$ evolves according to an Ornstein--Uhlenbeck (OU) process around $x_0\neq 0$:
\begin{equation}
\dot Y(t)
=
-[Y(t)-x_0]/\tau
+
\sqrt{2\sigma^2/\tau}\,\eta_Y(t).
\label{eq:Ylangevin}
\end{equation}
The Gaussian white noises are independent and satisfy $\langle\eta_\alpha(t)\rangle=0$ and $
\langle\eta_\alpha(t)\eta_\beta(t')\rangle
=
\delta_{\alpha\beta}\delta(t-t')$, 
where $\alpha,\beta\in\{X,Y\}$, $\delta_{\alpha\beta}$ is the Kronecker delta, and $\delta(t)$ is the Dirac delta function. The stationary OU process thus has mean $x_0$, variance $\sigma^2$, and correlation function
$\left\langle [Y(t)-x_0][Y(0)-x_0]\right\rangle
=
\sigma^2 e^{-|t|/\tau}$.
We choose the initial condition
$X(0)=0$ and $Y(0)=x_0$. The main quantity of interest is the capture time, defined as
\begin{equation}
T=\inf\{t>0:X(t)=Y(t)\}.
\label{eq:Tdef}
\end{equation}
Notably, if the resetting jump intersects with the target's position it is  not considered an encounter; this occurs only when the continuously evolving trajectories meet.

We now define the survival probability up to time $t$ as
$Q(x,y,t)=
{\Pr}[T>t\mid X(0)=x,Y(0)=y]$. Following the standard backward approach for diffusion with stochastic
resetting~\cite{EvansMajumdar2011}, and combining it with the OU process, the partial differential equation (PDE) for the survival probability up to time $t$ satisfies
\begin{align}
\partial_t Q
={}&D\,\partial_x^2 Q
+\frac{\sigma^2}{\tau}\,\partial_y^2 Q
-\frac{y-x_0}{\tau}\,\partial_y Q
\nonumber\\
&+r\left[Q(0,y,t)-Q(x,y,t)\right].
\label{eq:Qback}
\end{align}
Here, the first term describes the searcher's diffusion, while the
second and third terms are the diffusion and drift of the OU
target, respectively. The nonlocal term accounts for resetting: with rate $r$
the searcher is transferred from $x$ to the origin, while the target
coordinate remains unchanged. %This is the origin of the nonlocal
%contribution $Q(0,y,t)-Q(x,y,t)$.
Absorption occurs when the two meet; i.e., $Q(x,x,t)=0$,
with $Q(x,y,0)=1$ for $x\neq y$.

The capture time density equals
$-\dot Q(x,y,t)$. Thus, the MCT reads 
$M(x,y)
\!=\!-\!\int_0^\infty \!\!t\,\dot Q(x,y,t)dt\!=\!\int_0^\infty \!\!Q(x,y,t)dt$,
where the second equality follows by integration by parts.
As a result,  $M(x,y)$ 
satisfies the following PDE
\begin{align}
-\!1
\!=\!D \partial_x^2 M
\!+\!\frac{\sigma^2}{\tau}\partial_y^2 M
\!-\!\frac{y\!-\!x_0}{\tau}\partial_yM\!+\!
r\left[M(0,y)\!\!-\!\!M(x,y)\right]\!\!,
\label{eq:Mback}
\end{align}
with $M(x,x)=0$. The observable of interest is the MCT $\langle T\rangle=M(0,x_0)$, starting at $(0,x_0)$.

The central difficulty is apparent already from Eq.~\eqref{eq:Mback}. The absorbing boundary lies on the diagonal $x=y$, whereas resetting couples the solution nonlocally to the line $x=0$. Moreover, resetting acts only on the searcher and therefore does not renew the full state of the process. In particular, the target retains its position across reset events, generating correlations between successive search excursions. Consequently, the usual renewal description of resetting problems does not close. We therefore treat Eq.~\eqref{eq:Mback} numerically, except in the two limits where a separation of time scales allows an analytical treatment.

It is convenient to rescale the variables and parameters according to $z\!=\!x_0\sqrt{r/D}$, $\gamma\!=\!\sigma/x_0$, $\rho\!=\!r\tau$, $\bar{x}\!=\!x\sqrt{r/D}$ and
$\bar{y}\!=\!(y\!-\!x_0)/\sigma$,
and define $m(\bar{x},\bar{y})=M(x,y)/\tau$. As a result, the
backward PDE [Eq.~(\ref{eq:Mback})] becomes
\begin{equation}
-1
=
\rho\,\partial_{\bar{x}^2}m
+
\partial_{\bar{y}^2}m
-
\bar{y}\partial_{\bar{y}}m
+
\rho\left[m(0,\bar{y})-m(\bar{x},\bar{y})\right].
\label{eq:mback}
\end{equation}
The initial condition is now \(\bar{x}=0,\ \bar{y}=0\) for the searcher and target.
Moreover, the absorbing boundary condition on \(x=y\) becomes: $m\left(z(1+\gamma \bar{y}),\bar{y}\right)=0$.
The physical MCT satisfies
$\langle T\rangle
=\tau m(0,0)
=(\rho/r)\, m(0,0)$.

\textit{Slow target: \(\rho\to\infty\).}
In the limit of a slowly-moving target, the MCT can be written in a perturbative form
\begin{equation}
m(\bar{x},\bar{y})
=
\rho^{-1}m_0(\bar{x},\bar{y})
+
\rho^{-2}m_1(\bar{x},\bar{y})
+
O\left(\rho^{-3}\right).
\label{eq:mexp}
\end{equation}
In the leading order, the target is effectively frozen and \(m_0\) obeys
the standard resetting problem. For \(1+\gamma \bar{y}>0\) and
\(\bar{x}<z(1+\gamma \bar{y})\), one finds that
$m_0(\bar{x},\bar{y})
=
e^{z(1+\gamma \bar{y})}-e^{\bar{x}}$.
Hence, for the initial condition \(\bar{x}=0,\ \bar{y}=0\),
$m_0(0,0)
=e^z-1=rT_\infty$, 
where
$T_\infty\equiv (e^z-1)/r$ is the fixed-target MCT~\cite{EvansMajumdar2011}.
At sub-leading order, using~\eqref{eq:mback} one obtains
\begin{equation}
\partial_{\bar{x}^2}m_1
-
m_1(\bar{x},\bar{y})
+
m_1(0,\bar{y})
=
-\partial_{\bar{y}^2}m_0
+
\bar{y}\partial_{\bar{y}}m_0.
\label{eq:m1eq}
\end{equation}
Using the above result for 
$m_0(\bar{x},\bar{y})$,
we have $\partial_{\bar{y}}m_0 = \gamma z\,e^{z(1+\gamma \bar{y})}$ and $\partial_{\bar{y}^2}m_0 = \gamma^2z^2e^{z(1+\gamma \bar{y})}$, and the right hand side of Eq.~(\ref{eq:m1eq}) becomes $-e^{z(1+\gamma \bar{y})}
\left[
\gamma^2z^2-\gamma z\bar{y}
\right]$.
Therefore, for fixed \(\bar{y}\), Eq.~(\ref{eq:m1eq}) is a linear equation in \(\bar{x}\). Demanding that the solution does not diverge at \(\bar{x}\!\to\!-\infty\), together with the condition $m_1\left(z(1+\gamma \bar{y}),\bar{y}\right)=0$
yields
$m_1(\bar{x},\bar{y})
=
e^{z(1+\gamma \bar{y})}
\left[
e^{z(1+\gamma \bar{y})}-e^{\bar{x}}
\right]
\left[
\gamma^2z^2-\gamma z\bar{y}
\right]$.
Evaluating at the initial target position \(\bar{y}=0\), gives
$m_1(0,0)
= r  T_\infty\gamma^2z^2e^z$.
Therefore, defining the dimensionless MCT  $\mathcal{T}
= r\langle T\rangle$, we find
\begin{equation}
\mathcal{T}
=
\rho \,m(0,0)\simeq
\mathcal{T}_\infty\left(1+
\mathcal{A}/\rho\right),
\qquad
\mathcal{A}
=\gamma^2z^2e^z,
\label{eq:largeTau}
\end{equation}
where $\mathcal{T}_\infty=rT_\infty$. Thus, finite target mobility always increases the MCT
relative to the frozen-target limit at this order. In particular, $d\mathcal{T}/d\rho \simeq
-\mathcal{A}\mathcal{T}_\infty/\rho^2<0$, 
so the large-\(\rho\) branch is necessarily decreasing.

\textit{Fast target: \(\rho\to0\).}
The opposite limit is governed by the rapid relaxation of the target.
Introducing the fast time \(s=t/\tau\), the rescaled target's position $\bar{Y}$ satisfies
$d\bar{Y}/ds
=
-\bar{Y}+\sqrt{2}\,\eta_{\bar{Y}}(s)$.
On the fast time scale \(s=t/\tau\), a time interval \(ds\) corresponds
to \(dt=\tau ds\). Hence, the searcher variance grows as $\left\langle(\Delta \bar{x})^2\right\rangle
= 2\rho\,ds$,
while the reset probability is $\rho\,ds$.
Thus, both its effective diffusivity and reset rate  vanish as \(\rho\to0\). The searcher is therefore
effectively immobile at the origin, and the encounter problem reduces
to the first passage of the dimensionless OU target from \(\bar{y}=0\) to
\(\bar{y}=-1/\gamma\)~\cite{assaf2013extrinsic,assaf2013cooperation}.

Let \(u(\bar{y})\) denote the corresponding mean first-passage time in the
fast variable \(s\). It obeys $u''(\bar{y})-\bar{y}u'(\bar{y})=-1$ with $u\left(-1/\gamma\right)=0$.
At the initial point,
\begin{equation}\label{Phi}
u(0)
=
\sqrt{\pi}
\int_0^{1/(\sqrt{2}\gamma)}
e^{v^2}
\left[
1+\operatorname{erf}(v)
\right]
\,dv\equiv \Phi(\gamma).
\end{equation}
As a result, for $\rho\to 0$ we obtain $m(0,0) \simeq
\Phi(\gamma)$, and
\begin{equation}
\mathcal{T} \simeq
\rho\, \Phi(\gamma)
\label{eq:smallTau}
\end{equation}
Here, when $\gamma$ is small, the MCT is exponentially large in $1/\gamma^2$, whereas for $\gamma\gg 1$, $\mathcal{T}\sim \gamma^{-1}$.

\begin{figure}[t]
\includegraphics[width=0.88\columnwidth]{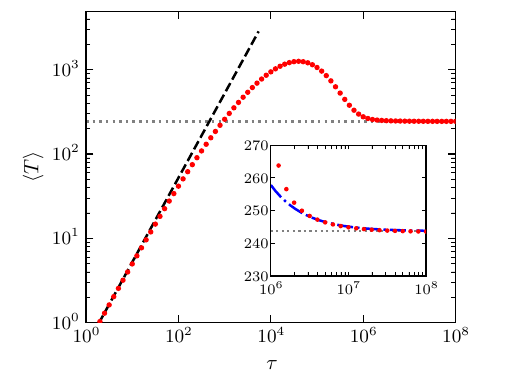}
\vspace{-5mm}\caption{
MCT vs. target correlation time for
$z\!=\!5.5$ and $\gamma\!=\!2.8$, with $D\!=\!r\!=\!1$.  Circles are the numerical solution of
Eq.~\eqref{eq:mback}. Dashed and dotted lines are the fast-target asymptote [Eq.~\eqref{eq:smallTau}], and frozen-target
value $T_\infty$. The dash-dotted curve in the inset is
the slow-target asymptote [Eq.~\eqref{eq:largeTau}].
}
\label{fig2}
\end{figure}

Equations~\eqref{eq:largeTau} and \eqref{eq:smallTau} already determine the qualitative structure of the problem. At small $\tau$ (or $\rho$), the MCT grows linearly, whereas at large $\tau$ (or $\rho$) it decreases toward $T_\infty$, giving rise to an intermediate maximum. At intermediate $\tau$, the target has sufficient mobility to wander far from the reset point but sufficient persistence to remain in an unfavorable configuration. Such excursions can significantly increase the MCT and generate the maximum. This is because the MCT is essentially determined by rare events where the excursion of the target is in the direction away from the searcher, which exponentially increases the MCT. 
Notably, the observed maximum is not associated with a local property of either limiting process, but with the competition between the fast-target and quasi-static regimes.
Figure~\ref{fig2} compares the numerical solution for $\langle T\rangle$ as a function of $\tau$ with the two asymptotic predictions. Good agreement is observed in the appropriate limits. Our numerical scheme discretizes Eq.~\eqref{eq:mback} on the $(\bar{x},\bar{y})$ plane, with absorption on $\bar{x}=z(1+\gamma\bar{y})$ and resetting coupled to $\bar{x}=0$. The resulting sparse linear system is solved for $m$ by the direct sparse solver SuperLU~\cite{Li2005}.

It is important to note that, increasing the fluctuation amplitude $\sigma$ can either facilitate or hinder encounters depending on $\tau$: for small $\tau$ stronger fluctuations promote rapid target-mediated encounters, whereas for large $\tau$ they increase the MCT.
%The agreement between  of the theoretical and numerical trends in the two limits suggests that the maximum results from the crossover between these two distinct dynamical mechanisms.
Another interesting consequence of target mobility appears in the limit of rapid resetting. For a fixed target, the MCT diverges exponentially as $r\to\infty$, as frequent resetting suppresses excursions of the searcher to the target. This divergence is regularized when the target is mobile. Indeed, for sufficiently large $r$ the searcher is effectively pinned at the reset position, $\bar{x}\simeq0$, while the OU target can still reach the origin by its own fluctuations. In this limit, the physical MCT satisfies $\langle T\rangle \simeq\tau \Phi(\gamma)$ [see  Eq.~(\ref{Phi})]. That is, $\langle T\rangle$ is finite and is independent on $r$; i.e., target mobility  converts the large-$r$ divergence
of the standard resetting problem into a finite plateau. 
%Our numerical
%solution confirms that the finite-$r$ minimum survives, while the
%high-reset branch approaches the target-controlled FP time rather than diverging.

\begin{figure}[t]
\includegraphics[width=\columnwidth]{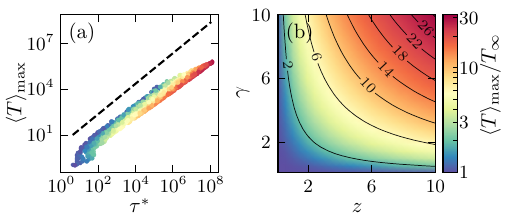}
\vspace{-9mm}\caption{
The maximum MCT $\langle T\rangle_{\text{max}}$ vs. the optimal correlation time $\tau_\star$, for $1024$ pairs $(z,\gamma)$ drawn uniformly at random from $[0.1,10]^2$ with $D=r=1$, so that $\rho=\tau$. The color denotes the ratio $\langle T\rangle_{\text{max}}/T_{\infty}$, which measures how strongly target mobility increases the MCT. The dashed line is a guide to the eye with unit slope. Right panel: the same ratio $\langle T\rangle_{\text{max}}/T_{\infty}$ across the $(z,\gamma)$
plane, on the same color scale.
}
\label{fig3}
\end{figure}

\textit{Crossover maximum estimate and optimal evasion strategy.}
Before we provide an estimation of the optimal correlation time $\tau_\star$ at which the MCT is maximized, we plot in
Fig.~\ref{fig3}  the numerical prediction of $\langle T\rangle_{\text{max}}$---the maximal MCT---versus the optimal correlation time $\tau_\star$.
The results approximately follow a linear scaling $\langle T\rangle_{\text{max}}\propto\tau_\star$ across many decades. 
The color coding shows that the enhancement $\langle T\rangle_{\text{max}}/T_\infty$ grows systematically with $z$ and $\gamma$; i.e., target mobility is most effective when the home position is far from the reset point and the explored range is wide. While the former is evident, the latter is counter-intuitive. Increasing $\sigma$ is expected to cause quicker encounters. However, it is the rare excursions of the target, in the oposite direction of the searcher, that control the MCT in this case.

Now, we provide a theoretical estimate of $\tau_\star$. As the maximum is generated by the crossover between the two limiting regimes, its location can be estimated by matching the corresponding asymptotics, Eqs.~(\ref{eq:largeTau}) and (\ref{eq:smallTau}):
$\Phi(\gamma)\rho_\star
=\mathcal{T}_\infty\left(1
+\mathcal{A}/\rho_\star\right)$.
The positive root is
\begin{equation}
\rho_\star
=[\mathcal{T}_\infty/(2\Phi(\gamma))]\left[1
+\sqrt{1+4\Phi(\gamma)\mathcal{A}/\mathcal{T}_\infty }\right]
.
\label{eq:tau1}
\end{equation}
For $\gamma\gtrsim{\cal O}(1)$, this result can be approximated in two limiting cases: 
\(\Phi\left(\gamma\right)\mathcal{A}/\mathcal{T}_\infty \ll 1\), which holds 
for $z\ll 1$ and the opposite limit, \(\Phi\left(\gamma\right)\mathcal{A}/\mathcal{T}_\infty \gg 1\), when $z\gg 1$. In the former limit, $\rho_\star\simeq \mathcal{T}_\infty/\Phi(\gamma)$, whereas in the latter limit, $\rho_\star
\simeq
\sqrt{\mathcal{T}_\infty \mathcal{A}
/\Phi(\gamma)}$. Importantly, Eq.~\eqref{eq:tau1} is a crossover estimate and thus should not be interpreted as an exact extremum condition. This is because the maximum is found   outside the region of applicability of both small-$\tau$ and large-$\tau$ expansions. As such, Eq.~(\ref{eq:tau1}) estimates the scale at which the two asymptotic mechanisms become comparable. Thus, upon comparison with numerics, an overall \(O(1)\) calibration \(\rho_\star\to C\rho_\star\) may be required, see below.  Incidentally, a similar matching procedure, using a crossover between two limiting regimes, has proven useful in finding the optimal correlation time of fluctuating environments in various settings~\cite{assaf2013extrinsic,assaf2013cooperation,assaf2017wkb,taitelbaum2020population}.

%The first contribution is the naive crossover between the fast-target law and the frozen-target result. The second is the leading correction arising from the finite mobility of the target. Thus, although the estimate is obtained by asymptotic matching, it explicitly separates the static resetting scale from the correction induced by target fluctuations. 

\begin{figure}[t]
\includegraphics[width=\columnwidth]{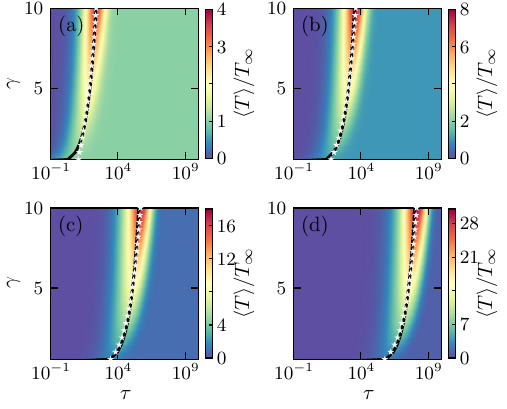}
\vspace{-9mm}
\caption{
Ratio $\langle T\rangle/T_\infty$ between the MCT and its frozen-target value,
as a function of the target correlation time $\tau$ and the relative fluctuation
amplitude $\gamma=\sigma/x_0$, for $D=r=1$ (so that $\rho=\tau$) and
(a) $z=1$, (b) $z=2.35$, (c) $z=5.5$, and (d) $z=10$. White stars mark the numerically determined optimal
correlation time $\tau_\star(\gamma)$. The solid curve is the crossover estimate
[Eq.~\eqref{eq:tau1}] with a fitted multiplicative factor of $3.5$, $4.9$, $9.2$, and $16.0$ in panels (a-d), respectively.
}
\label{fig4}
\end{figure}

Figure~\ref{fig4} shows the relative height of the maximal MCT compared with the fixed-target result, as function of the target's characteristics: correlation time and home range. The stars denote the numerical calculation of the optimal correlation time, which agrees well with the theoretical estimate  [Eq.~(\ref{eq:tau1})]  (solid line) with a calibration  factor $C(z)$, see caption~\footnote{
To corroborate the numerical solution of Eq.~(\ref{eq:mback}) we also ran Monte-Carlo simulations. Here, each trajectory follows the searcher at $X(t_n)$ and target at $Y(x_n)$ together on a time grid
$t_{n+1}=t_n+h$ with $h\le 0.05\,\tau$. %Resetting times are drawn as exponential waiting times of rate $r$.
%\begin{equation*}
%X(t_{n+1})=X(t_n)+\sqrt{2Dh}\,\xi_n ,
%\qquad
%Y(t_{n+1})=x_0+\bigl[Y(t_n)-x_0\bigr]e^{-h/\tau}
%+\sigma\sqrt{1-e^{-2h/\tau}}\,\eta_n ,
%\end{equation*}
%where $\xi_n$ and $\eta_n$ are independent standard Gaussian variables. 
Denoting by
$\Delta_n=Y(t_n)-X(t_n)$, a meeting is recorded whenever $\Delta_n\Delta_{n+1}\le 0$. 
%Otherwise, it is recorded with  probability $p_n=\exp\!\left[-\Delta_n\Delta_{n+1}/[(D+\sigma^2/\tau)\,h]\right]$ that a continuous random trajectory interpolation pinned between $\Delta_n$ to
%$\Delta_{n+1}$ touches zero, 
%accounting for crossings between grid points.
We ran a sufficient number of realizations 
%$2\times10^{4}$---$5\times10^{6}$ realizations 
for each data point, to keep a statistical error below $1\%$.}. The competition between the fast- and slow-target limits therefore governs optimal evasion across the full parameter range. Increasing $z$ or $\gamma$ shifts the optimal correlation time to larger values and strengthens the delay relative to a frozen target, indicating that persistence is most effective when the target's home is far from the reset point and its fluctuations are wide. The figure also confirms that the intermediate-$\tau$ maximum of Fig.~\ref{fig2} is a generic feature of the model.

%\textit{Numerics and discussion.}
%Equation~\eqref{eq:Mback} is discretized on the $(x,y)$ plane with absorption imposed on the diagonal $x=y$ and the reset term coupled explicitly to the grid line $x=0$. For the broad maxima, which occur at $\tau\gg1$ in the parameter ranges of Figs.~\ref{fig:rscan} and \ref{fig:sigmascan}, we use a slow-target numerical continuation. At fixed target position $y>0$, the stationary resetting searcher has mean encounter time
%\begin{equation}
%T_{\rm stat}(y)=\frac{e^{y\sqrt{r/D}}-1}{r},
%\end{equation}
%and corresponding inverse time scale
%\begin{equation}
%k(y)=\frac{r}{e^{y\sqrt{r/D}}-1}.
%\end{equation}
%The slow-target continuation is obtained by combining this local search rate with the OU backward operator. Its overlap with the direct two-dimensional solution is illustrated in Fig.~\ref{fig2}; it is numerically inexpensive and resolves the peak over wide scans in $r$ and $\sigma$.

%\textcolor{red}{[SR: To save space, I think it is safe to remove the  paragraph below. We already said these things and in PRL you have to minimize repetitions.]}The physical origin of the maximum is straightforward. When $\tau$ is very small, the target decorrelates rapidly and repeatedly samples positions near the origin, so the target itself efficiently produces the encounter. When $\tau$ is very large, the target is effectively stationary and resetting repeatedly launches independent search excursions toward it. 

Notably, the fact that the maximal region in Fig.~\ref{fig4} is narrow  with respect to $\tau$ indicates that 
if the target is a prey that wants to optimally evade from a predator, apart from increasing its fluctuation's magnitude, which may come at a cost, the prey can precisely tune the correlation time in a smart manner to maximize the MCT. Also note that the fixed-target MCT is minimized at $z\simeq 1.59$~\cite{EvansMajumdar2011}, which lies between panels~(a) and~(b) of Fig.~\ref{fig4}. Nonetheless, by countering this behavior and tuning the correlation time properly, the target prey can elongate the evasion essentially indefinitely. 

%The two asymptotic scales also reveal useful parameter dependences. The fast-target coefficient $\Phi$ depends only on $x_0/\sigma$, whereas the stationary-target result depends on the resetting combination
%$z=x_0\sqrt{r/D}$. Thus the crossover is governed by the competition between the geometrical fluctuation ratio $x_0/\sigma$ and the resetting length $\sqrt{D/r}$. The residual mismatch between the crossover estimates and the numerical maxima in Fig.~\ref{fig4} shows that no universal multiplicative conversion from crossover scale to exact peak position should be expected without a theory of the intermediate regime.

\textit{Conclusion.}  We studied the evasion of a mobile target from a resetting searcher. For a target fluctuating around a home location, we have identified an optimal correlation time that maximizes the mean capture time (MCT). Analytical results in the two limiting regimes establish the origin of this optimum: the MCT vanishes as the target correlation time tends to zero, and approaches the frozen-target value from above as it tends to infinity. Rapid target motion therefore facilitates capture, whereas sufficiently persistent excursions can hinder it, producing a maximum between these extremes.

Matching the asymptotic regimes provides a crossover estimate for the optimal correlation time $\tau_\star$, predicting an approximately linear relation between the maximal MCT and $\tau_\star$ across many decades. Wider target fluctuations can enhance the delay relative to a frozen target when their persistence is appropriately tuned, highlighting the benefit of excursions away from the searcher's reset point. Target mobility also regularizes divergent MCTs caused by rapid resetting: when the searcher is effectively pinned at the origin, capture is controlled by the target's own first passage to that point.

This work shifts the focus of resetting-based search from expediting discovery to delaying capture. Our results identify temporal correlations as a resource for evasion: even at a fixed stationary spatial distribution, a target can substantially prolong its mean survival by tuning the persistence of its motion. Successful evasion thus depends not only on where a target moves, but also on how its motion unfolds in time.

\textit{Acknowledgments. }AT and MA acknowledge support from ISF grant 2084/26. SR acknowledges support from ERC under the European Union’s Horizon 2020 research and innovation program (Grant Agreement No. 947731).

%Extensions to higher dimensions, and resetting of both coordinates should provide further examples in which memory and restart compete to control search efficiency. It would also be interesting to investigate the distribution itself of encounter times rather than only the mean.

%\textbf{Additional issues:} (i) Is there a phase transition with respect to the position of the optimal $r$? does it change from finite optimal $r$ to $0$ or $\infty$? Possibly when the distribution of waiting times between resets is changed from exponential to a different distribution. (ii) Resets of both target and searcher instead of OU decay. (iii) different distribution of reset times (other than exponential).

%\bibliographystyle{apsrev4-2}
\bibliography{bibtex}
\end{document}